\documentclass[12pt]{article}
\usepackage{amsmath}
\usepackage{amssymb}
\usepackage{amsfonts}
\usepackage{eucal}
\usepackage[usenames]{color}
\usepackage{graphicx}
\usepackage{bbold}
\usepackage{bbm}
\usepackage{stackrel}
\usepackage{cancel}
\usepackage{enumitem}

\numberwithin{equation}{section}
\newcommand{\E}{\mathbb{E}}

\DeclareMathOperator\erf{erf}
\DeclareMathOperator{\erfc}{erfc}

\begin{document}
\def\opsim{\mathop{\sim}}
\def\opsimeq{\mathop{\simeq}}
\title{First-passage time distribution of a Brownian motion: two unexpected journeys.}
 \author{Alain Mazzolo\thanks{Universit\'e Paris-Saclay, CEA, Service d'\'Etudes des R\'eacteurs et de Math\'ematiques Appliqu\'ees, 91191, Gif-sur-Yvette, France. E-mail: \texttt{alain.mazzolo@cea.fr}}  }
%%\vspace{10pt}

\date{}
\maketitle

\begin{abstract}
The distribution of the first-passage time (FPT)$T_a$ for a Brownian particle with drift $\mu$ subject to hitting an absorber at a level $a>0$ is well-known and given by its density $\gamma(t) = \frac{a}{\sqrt{2 \pi t^3} }  e^{-\frac{(a-\mu t)^2}{2 t}}, t>0$, which is normalized only if $\mu \geq 0$. This article demonstrates the existence of two additional diffusion process categories (one with one parameter and the other with two) that have the same first passage-time distributions when $\mu <0$. For both, we identify the transition densities and thoroughly investigate the processes. A substantial implication is that the first-passage time distribution does not indicate whether the process originates from a drifted Brownian motion or from one of the new processes presented.
\end{abstract}

%%\newtheoremstyle{thmstyle}
%%	{9pt}
%%	{9pt}
%%	{\itshape}
%%	{}
%%	{\bfseries\scshape}
%%	{.}
%%	{\newline}
%%	{}
%%\theoremstyle{thmstyle}
%%\newtheorem{thm}{Theorem}

%%\pacs{02.50.Ey, 05.10.Gg, 05.40.Jc}

%%\keywords{taboo process, constraint, stochastic differential equation, spectral gap}

%%%%%%%%%%%%%%%%%%%%%%%%%%%%%%%%%%%%%%%%%%%%%%%%%%%%%%%%%%%%%%%%%%%%%%%%%%%%%%%%%%%%%%%%%%%%%%%%%%%%%%%%%%%%%%%%
%%																						%%
%%                               INTRODUCTION                                                                 %%
%%																						%%
%%%%%%%%%%%%%%%%%%%%%%%%%%%%%%%%%%%%%%%%%%%%%%%%%%%%%%%%%%%%%%%%%%%%%%%%%%%%%%%%%%%%%%%%%%%%%%%%%%%%%%%%%%%%%%%%
\section{Introduction}
\label{sec_intro}
Finding the distribution for the first time when a diffusion reaches a boundary~\cite{ref_Darling,ref_book_Redner} is a fundamental task in characterizing the process and presents applications in multiple disciplines, including stochastic analysis~\cite{ref_Abundo_1,ref_Abundo_2,ref_Hernandez}, mathematical finance, physics, chemistry~\cite{ref_book_Metzler}, and biology~\cite{ref_Ricciardi_biology}, particularly in the context of animal movement~\cite{ref_McKenzie} and neuroscience~\cite{ref_Sacerdote}. Quantities directly related to first-passage time, such as first-passage Brownian functionals, have also been the subject of extensive research~\cite{ref_Majumdar_CS,ref_Majumdar-Meerson,ref_Abundo_Vescovo}.  However, despite substantial efforts, certain problems remain open, such as the first-passage time of the Ornstein-Uhlenbeck process, whose closed-form distribution is only known in the special case where the barrier is located at zero~\cite{ref_Alili,ref_Ricciardi}. \\
In this article, we begin with the drifted Brownian motion, which is defined by its stochastic representation:
\begin{equation}
\label{def_BM_with_drift_SDE}
   \left\{
       \begin{aligned}
	  dX(t) & = \mu dt + \sigma dW(t) \, , ~~ t \geq 0   \\       
	  X(0)  & = 0 \, ,
       \end{aligned}
   \right.
\end{equation}
\noindent where $W(t)$ is a standard Brownian motion (Wiener process), $\mu$ the constant drift of the process and $\sigma$, the positive diffusion coefficient (which does not play any role in our study), is set to $1$. Let us denote by $T_a$ the stopping time:
\begin{equation}
\label{def_stopping_time}
	T_a = \inf_{t \ge 0} \{t: X(t) = a \} \, ,
\end{equation}
where the barrier $a$, without loss of generality is positive. The first-passage time distribution $T_a$ of this Brownian motion can be obtained using the reflection principle and its density, with respect to the Lebesgue measure, is the well-known inverse Gaussian density given by~\cite{ref_book_Karatzas}
\begin{equation}
\label{distribution_Ta}
	\gamma(t) = \frac{a}{\sqrt{2 \pi  t^3} }  e^{-\frac{(a-\mu t)^2}{2  t}} \, , ~~ t>0,
\end{equation}
and is normalized to unity only if $\mu \geq 0$. More precisely,
\begin{eqnarray}
\int_{0}^{+\infty} \gamma(t) dt 
=  \left\lbrace
  \begin{array}{lll}
    1  
    &~~\mathrm{if~~} \mu \ge 0
    \\
    e^{-2 \vert \mu \vert a } 
    &~~\mathrm{if~~} \mu < 0  \, .
  \end{array}
\right.
\label{normalisation}
\end{eqnarray}
%Thus far, the first crossing density $\gamma(t)$ has always been associated with the drifted Brownian motion, a result that goes back to Schrodinger in 1915. This result raises an important question: are there others processes, different from the Brownian motion, that have the same the same density $\gamma(t)$ for the first crossing? Of course there is two possible answers: "yes" then we have to found out at least one process (different from the Brownian motion) that has the same density $\gamma(t)$, "no" then we have to prove that the density $\gamma(t)$ corresponds to only one "canonical process" (here the Brownian motion).
%The purpose of this article is demonstrate that the correct answer is "yes", by showing that there are specific diffusion type processes, different from the "canonical", and such that they have the same density $\gamma(t)$ for the first crossing.
%% inverse Gaussian distribution
Thus far, the first passage-time density $\gamma(t)$ has always been associated with the drifted Brownian motion, a result that goes back to Schr\"odinger in 1915~\cite{ref_Schrodinger}. This finding raises an important question: are there processes other than drifted Brownian motion, that have the same first passage-time density $\gamma(t)$? There are, of course, two possible answers to this question: 
\begin{itemize}
\item "Yes" then we have to find at least one process, distinct from drifted Brownian motion, that shares the same density $\gamma(t)$."
\item "No" then we have to prove that the density $\gamma(t)$ corresponds to only one "canonical process" (here the drifted Brownian motion).
\end{itemize}

\noindent The purpose of this article is to demonstrate that the correct answer is "Yes", by presenting specific diffusion-type processes that are distinct from the drifted Brownian motion and share the same density $\gamma(t)$ for the first passage-time density.\\

\noindent To this aim, we will constrain the drifted Brownian motion in such a way that it has a new first-passage time distribution, which is again given by the relation\eqref{distribution_Ta}, but with a distinct parameter $\lambda \neq \mu$. In the recent publication~\cite{ref_Monthus-Mazzolo}, we found out that if $\mu \geq 0$ and $\lambda \geq 0$ then the constrained process is merely a drifted Brownian motion with drift $\lambda$, which is somewhat as expected. In this article, we focus on the scenario where $\lambda < 0$, and depending on the sign of $\mu$, we obtain remarkably unexpected results, 
%% which are in fact far from a drifted Brownian 
since the drifts of the conditioned processes are both time- and space-dependent, proving the existence of diffusion processes that possess identical first passage-time density $\gamma(t)$, as the drifted Brownian motion.\\

%In order to clarify the situation, the paper is organized as follows: in Section~\ref{sec2} we recall what one needs to know about conditioning with respect to the first-passage time in order to derive the effective Langevin equations satisfied by the conditioned stochastic processes. Then, in section~\ref{sec3} we apply the formalism when the new distribution of the first-passage time is given by relation\eqref{distribution_Ta} with a negative parameter. In section~\ref{sec4} we solve analytically the effective stochastic Langevin equations using Girsanov's theorem and we study the densities obtained. Finally, Section~\ref{sec_Discussion} presents some concluding remarks and open questions. \ref{appendix_1} and~\ref{appendix_2} provide calculation details.

The paper is structured as follows: Section~\ref{sec2} provides a summary of the information required for conditioning with respect to the first-passage time and deriving the conditioned stochastic differential equations for conditioned stochastic processes. Section~\ref{sec3} applies this formalism to cases where the new distribution of the first-passage time is defined by relation\eqref{distribution_Ta} with a negative parameter. In Section~\ref{sec4}, we analytically solve the conditioned stochastic differential equations by utilizing Girsanov's theorem and study the resulting densities. Finally, Section~\ref{sec_Discussion} presents some concluding remarks and open questions. Calculation details are given in Appendices~\ref{appendix_1} and~\ref{appendix_2}.
%%%%%%%%%%%%%%%%%%%%%%%%%%%%%%%%%%%%%%%%%%%%%%%%%%%%%%%%%%%%%%%%%%%%%%%%%%%%%%%%%%%%%%%%%%
%%																						%%
%%            Conditioning with respect to the first-passage time, a brief remainder    %%
%%																						%%
%%%%%%%%%%%%%%%%%%%%%%%%%%%%%%%%%%%%%%%%%%%%%%%%%%%%%%%%%%%%%%%%%%%%%%%%%%%%%%%%%%%%%%%%%%
\section{Conditioning with respect to the first-passage time, a brief remainder}
\label{sec2}

%%%%%%%%%%%%%%%%%%%%%%%%%%%%%%%%%%%%%%%%%%%%%%%%
%%       SUBSECTION: GENERAL SETTING           %
%%%%%%%%%%%%%%%%%%%%%%%%%%%%%%%%%%%%%%%%%%%%%%%%
%%\subsection{General setting}
In this section, we recall the essential ingredients to derive the stochastic differential equation (SDE) for a diffusion conditioned by its first-passage time. This type of conditioning is a special case of conditioning with respect to a random time, whose general theory is still missing. To the best of our knowledge, the first work in this field is due to Baudoin~\cite{ref_Baudoin} and was continued in~\cite{ref_Larmier}, but it is only very recently that conditioning with respect to first-passage time and the survival probability, which is the
probability that a particle did not get absorbed at the boundary up to time $t$, has been established~\cite{ref_Monthus-Mazzolo}. We refer the readers to this article for a precise meaning of conditioning with respect to a density probability and an event.\\
%% Regarding the conditioning with respect to the survival probability, see also the recent article~\cite{ref_Pozzoli} which investigates the transport properties of diffusive particles conditioned to survive in trapping environments.\\
In the following, we consider a Brownian motion with constant drift $\mu$ with absorbing condition at a position $a>0$.  
The unconditioned process $X(t)$ has the stochastic representation~\eqref{def_BM_with_drift_SDE}, whose  transition density $P_{\mu}(x,t\vert 0,0) = P_{\mu}(x,t)$ can be obtained by the method of images~\cite{ref_book_Grimmett}
\begin{equation}
\label{imagesBrown}
	P_{\mu}(x,t)  = \frac{1}{\sqrt{2 \pi t}} \left( e^{- \frac{(x-\mu t )^2}{2t}} 
- e^{ 2 \mu a } e^{- \frac{(x-2a-\mu t)^2}{2t}} \right) , ~~ t>0 \,.
\end{equation}
With the transition density available, the distribution of the first-passage time follows easily~\cite{ref_book_Redner,ref_Monthus-Mazzolo}, its density is
\begin{equation}
\label{distribution_Ta_bis}
	\gamma(t) = \frac{a}{\sqrt{2 \pi  t^3} }  e^{-\frac{(a-\mu t)^2}{2  t}}  , ~~ t>0 \, ,
\end{equation}
as well as the survival probability $S_{\mu}(t)$, which is the probability that the drifted Brownian motion has not hit the absorbing boundary at time $t$. This quantity can be calculated in two ways, using either the transition density or the first-passage time distribution~\cite{ref_book_Redner}
 
\begin{align}
\label{survival_at_time_t}
	S_{\mu}(t) & = 1 - \int_0^t \gamma(s) ds = \int_{-\infty}^a P_{\mu}(x,t) dx\nonumber \\
%	           & = \frac{1}{2} \left(1 + \text{erf}\left(\frac{a-\mu  t}{\sqrt{2 t}}\right) - e^{2 a \mu } \text{erfc}\left(\frac{a+\mu  t}{\sqrt{2 t}}\right)\right) \\
	           & =  \Phi\left(a/\sqrt{t} - \mu \sqrt{t} \right)  + e^{2 a \mu } \left(1 - \Phi\left(a/\sqrt{t} + \mu \sqrt{t} \right) \right) \,
\end{align}
which is the well-known Bachelier-L\'evy formula, where 
\begin{equation}
\label{def_distribution_function_gaussian}
	\Phi(x) = \frac{1}{\sqrt{2 \pi}} \int_{-\infty}^x e^{-u^2/2} du 
\end{equation}
is the distribution function of the standard Gaussian random variable.
%%where $\erf(x) = \frac{2}{\sqrt{\pi}} \int_0^{x} du \, e^{- u^2}$ is the Error function and $\erfc(x)$ the complementary Error function $\erfc(x) = 1 - \erf(x)$. 
In particular, the forever-survival probability $ S_{\mu}(\infty)$ vanished only for positive drift $\mu \geq 0$
\begin{eqnarray}
S_{\mu}(\infty ) =  
 \left\lbrace
  \begin{array}{lll}
    0  
    &~~\mathrm{if~~} \mu \ge 0
    \\
    1- e^{-2 \vert \mu \vert a }
    &~~\mathrm{if~~} \mu < 0  \,  ,
  \end{array}
\right.  
\label{survivalBrown}
\end{eqnarray}
while for negative drift $\mu<0$, the particle can escape toward $(-\infty)$ without touching the absorbing boundary $a$ with the finite probability $S_{\mu}(\infty)= 1- e^{-2 \vert \mu \vert a } $. Now, we want to impose a new probability distribution for the first-passage time. From~\cite{ref_Monthus-Mazzolo} we know that when the time horizon, i.e. the time when the survival is imposed, is infinite $T=+\infty$, the conditioning consists in imposing the new probability distribution with density $\gamma^*(t)$ of the absorption time $T_a \in [0,+\infty[$, whose normalization $[1- S^*(\infty )]$ determines the conditioned probability $S^*(\infty ) \in [0,1]$ of the forever-survival probability. Thus, there are two linked conditions, the first concerning the probability distribution of the absorption time and the second the survival probability. The conditioned stochastic differential equation~\cite{ref_Baudoin} for the conditioned process $X^*(t)$ is described by the following SDE:
%%The effective Langevin equation for the conditioned process $X^*(t)$ is described by the following SDE:\footnote{In physics literature, this type of SDE is often referred as an effective Langevin equation~\cite{ref_Majumdar_Orland}, while in the field of mathematics it is called a conditioned stochastic differential equation~\cite{ref_Baudoin}.}
\begin{equation}
\label{Ito_condioned}
	X^*(t) = \mu^*(X^*(t),t)dt + dW(t)  \,  ,~~ t \geq 0 ,
\end{equation}
where the conditioned drift $\mu^*(x,t)$ is given by
\begin{equation}
	\mu^*(x,t)  = \mu + \partial_x \ln Q(x,t) \, ,
\end{equation}
with the function 
%%\begin{eqnarray}
%% \label{Qdef}
%%	Q(x,t) &= &\int_t^{\infty} \gamma^*(s) \frac{\gamma(s \vert x,t)}{\gamma(s\vert x_0,0)}  ds \nonumber \\
%%	       & +&S^*(\infty ) \left[ \lim_{T \to +\infty} 
%%\frac{ S(\infty \vert x) +  \int_{T}^{+\infty}  \gamma(s \vert x,t) ds }
%% {S(\infty \vert x_0) +  \int_{T}^{+\infty}  \gamma(s \vert x_0,0) ds }
%% \right]  \, ,
%%\end{eqnarray}
\begin{equation}
 \label{Qdef}
	Q(x,t) = \int_t^{\infty} \gamma^*(s) \frac{\gamma(s \vert x,t)}{\gamma(s\vert x_0,0)}  ds 
	       +S^*(\infty ) \left[ \lim_{T \to +\infty} 
\frac{ S(\infty \vert x) +  \int_{T}^{+\infty}  \gamma(s \vert x,t) ds }
 {S(\infty \vert x_0) +  \int_{T}^{+\infty}  \gamma(s \vert x_0,0) ds }
 \right]  \, ,
\end{equation}
containing all the information concerning the conditioning~\cite{ref_Monthus-Mazzolo}. This expression can be viewed as a sophisticated Doob h-transform~\cite{ref_Doob,ref_book_Doob,ref_book_Rogers}. In the previous equation, $S(\infty \vert x)$ is the survival probability when starting at position $x$, $\gamma(t \vert x,t)$ is the conditional probability density of the absorption-time $T_a$ knowing that the process started at the position $x$ at time $t$, and $x_0 = X(0)$. Additionally, from its construction, $Q(x,t)$ inherits the backward Fokker-Planck dynamics of the initial process with respect to the initial variables $(x,t)$, that is
\begin{equation}
\label{Qbackward}
	- \partial_t  Q(x,t)  =     \mu \, \partial_{x} Q(x,t)  + \frac{1}{2}\partial^2_{x x} Q(x,t)  \, .
\end{equation}
For more details, again we refer the reader to the paper~\cite{ref_Monthus-Mazzolo} which provides proofs and numerous examples. In particular, when 
\begin{equation}
\label{distrib_final}
	\gamma^*(t)  =  \frac{a }{\sqrt{2 \pi t^3} } e^{- \frac{(a-\lambda t )^2}{2 t}}, ~~ t > 0  \mathrm{~~with~} \lambda ~\geq 0  \, , 
\end{equation}
and thus $S^*(\infty ) = 0$, the conditioned drift $\mu^*(x,t)$ reduces to $\lambda$, which is the expected result. From now on, we consider that the target first-passage density $\gamma^*(t)$ is given by Eq.\eqref{distrib_final} with $\lambda < 0$ and we distinguish two cases based on whether the initial drift $\mu$ is positive or strictly negative.

%%%%%%%%%%%%%%%%%%%%%%%%%%%%%%%%%%%%%%%%%%%%%%%%%%%%%%%%%%%%%%%%%%%%%%%%%%%%%%%%%%%%%%%%%%%%%%%%
%%	                                                                                      %%
%%   Conditioning with respect to the first-passage time:                                     %%  
%%   $\gamma^*(T_a )  =  \frac{a }{\sqrt{2 \pi T_a^3} } e^{- \frac{(a-\lambda T_a )^2}{2T_a}} %%
%%   $ with $\lambda < 0$                                                                     %%
%%	                                                                                      %%
%%%%%%%%%%%%%%%%%%%%%%%%%%%%%%%%%%%%%%%%%%%%%%%%%%%%%%%%%%%%%%%%%%%%%%%%%%%%%%%%%%%%%%%%%%%%%%%%
\section{Conditioning with respect to the first-passage time density $\gamma^*(t)  =  \frac{a }{\sqrt{2 \pi t^3} } e^{- \frac{(a-\lambda t )^2}{2 t}} $ with $\lambda < 0$}
\label{sec3}
When conditioning toward a distribution of first-passage time whose density is not normalized to unity, the forever survival probability also plays a crucial role as established in~\cite{ref_Monthus-Mazzolo}.
The expression of $Q(x,t)$ given by Eq.\eqref{Qdef} must be computed carefully and is strongly dependent on the sign of the initial drift $\mu$. Therefore, in the two upcoming subsections, we will distinguish the two cases, based on whether the initial drift is zero or positive $\mu\geq 0$ or strictly negative $\mu < 0$.

%%%%%%%%%%%%%%%%%%%%%%%%%%%%%%%%%%%%%%%%%%%%%%%
%%       SUBSECTION: POSITIVE DRIFT  mu >= 0  %
%%%%%%%%%%%%%%%%%%%%%%%%%%%%%%%%%%%%%%%%%%%%%%%
\subsection{ When the initial drift is vanishing or positive $\mu\geq 0$ (case I) } 
When the initial drift is $\mu\geq 0$, the conditional probability density $\gamma(s\vert x,t)$ of the absorption at time $T_a$ knowing that the process started at the position $x$ at time $t$ reads~\cite{ref_book_Redner,ref_Monthus-Mazzolo}
\begin{equation}
\label{gammafirstBrown}
	\gamma(s \vert x,t) =  \frac{(a-x) }{\sqrt{2 \pi (s-t)^3} } e^{- \frac{(a-x-\mu(s-t) )^2}{2(s-t)}}  \, , ~ s>t \geq 0,
\end{equation}
and the forever survival probability vanished $S(\infty \vert x)=S(\infty \vert 0)=0$. Moreover, since 
\begin{equation}
\label{gammaconditionedBrown}
	\gamma^*(t)  =  \frac{a }{\sqrt{2 \pi t^3} } e^{- \frac{(a-\lambda t)^2}{2 t}}  , ~~ t>0 \, ,
\end{equation}
with $\lambda < 0$, the forever survival probability associated is $ S_{\lambda}^*(\infty) = (1- e^{2 \lambda a}) $. Reporting these expressions in~Eq.\eqref{Qdef}, we get
\begin{align}
\label{Qmuposlambda}
 Q_{I}(x,t) 
& =   \left( \frac{a-x}{a} \right)  e^{ - \mu x  + \frac{\mu^2 }{2} t} 
\left[ \int_t^{+\infty}   
\frac{a }{\sqrt{2 \pi s^3} } e^{\lambda a - \frac{a^2}{2 s}- \frac{\lambda^2}{2} s} 
  \left(\frac{s}{s-t}  \right)^{\frac{3}{2} }   
e^{\frac{a^2}{2 s} - \frac{(a-x)^2}{2(s-t)}} ds
+  (  1- e^{2 \lambda a } )   \right]
\nonumber \\
%% & =   \left( \frac{a-x}{a} \right)  e^{ - \mu x  + \frac{\mu^2  }{2} t} 
%% \left[ \frac{a }{\sqrt{2 \pi}  }e^{\lambda a - \frac{\lambda^2  }{2} t}
%% \int_t^{+\infty}    
%%   \left(\frac{1}{s-t}  \right)^{\frac{3}{2} }   
%% e^{ - \frac{\lambda^2}{2}(s-t) - \frac{(a-x)^2}{2(s-t)}} ds
%% +  (  1- e^{2 \lambda a } )   \right]
%% \nonumber \\
%% & =   \left( \frac{a-x}{a} \right)  e^{ - \mu x  + \frac{\mu^2  }{2} t} 
%% \left[ \frac{a }{\sqrt{2 \pi}  }e^{\lambda a - \frac{\lambda^2  }{2} t}
%% \frac{ \sqrt{2 \pi }}{ (a-x)} e^{  \lambda  (a-x) }
%% +  (  1- e^{2 \lambda a } )   \right]
%% \nonumber \\
& =    e^{ - \mu x  + \frac{\mu^2  }{2} t + 2 \lambda a} 
\left[ e^{  - \lambda  x - \frac{\lambda^2  }{2} t}
+  (   e^{- 2 \lambda a } -1) \left( \frac{a-x}{a} \right)  \right]   \, ,
\end{align}
and the corresponding drift
\begin{align}
\label{driftdoobmuposlambda}
 \mu^*_{I}(x,t) &  = \mu +  \partial_x \ln  Q_{I}(x,t)  \nonumber  \\
			&  =  \partial_x \ln 
\underbrace{\left[ e^{  - \lambda x - \frac{\lambda^2}{2}t } + (   e^{-2 \lambda a } -1)   \left(  \frac{a-x }{a}   \right) \right]}_{\equiv  \tilde{Q}_{I}(x,t) }
\nonumber \\
            & =  
\frac{ e^{  - \lambda x - \frac{\lambda^2}{2}t } (-\lambda) + (   e^{-2 \lambda a } -1)   \left( - \frac{1 }{a}   \right) }
{ e^{  - \lambda x - \frac{\lambda^2}{2}t } + (   e^{-2 \lambda a } -1)   \left(  \frac{a-x }{a} \right)   } \, .
\end{align} 
Consequently, the stochastic representation of the conditioned process, denoted as $X_I(t)$, is not that of a Brownian motion with drift, $dX_I(t) = \lambda dt + dW(t)$ (as one might have expected) but obeys an inhomogeneous diffusion process
\begin{align}
\label{SDEdriftdoobmuposlambda}
   dX_I(t) = \left(  \frac{ e^{  - \lambda X_I(t) - \frac{\lambda^2}{2}t } (-\lambda) + (   e^{-2 \lambda a } -1)   \left( - \frac{1 }{a}   \right) }
{ e^{  - \lambda X_I(t) - \frac{\lambda^2}{2}t } + (   e^{-2 \lambda a } -1)   \left(  \frac{a-X_I(t) }{a} \right)   }    \right) dt + dW(t)  \, , ~~ t \geq 0 \,  , 
\end{align} 
which is independent of the initial drift $\mu$.  
Assuming that $X_I(t)$ remains finite, the long-time behavior of the conditioned drift 
\begin{equation}
\label{logtimemuI}
	\mu^*_{I}(x,t)   \underset{t \to \infty}{\sim\,} -\frac{1}{a-x}
\end{equation}
corresponds to the drift of a taboo process (with taboo state $a$)~\cite{ref_Knight,ref_Pinsky,ref_Mazzolo_Taboo}. In particular, the  boundary $a$ is no longer absorbing, but becomes a entrance boundary~\cite{ref_book_Karlin}, which means that the boundary $a$ cannot be reached from the interior of the state space, i.e. the interval $]-\infty,a[$. Roughly speaking, after an arbitrarily large time $t$, the process can no longer be absorbed, which is the expected behavior.

%%%%%%%%%%%%%%%%%%%%%%%%%%%%%%%%%%%%%%%%%%%%%%%
%%       SUBSECTION: NEGATIVE DRIFT  mu < 0   %
%%%%%%%%%%%%%%%%%%%%%%%%%%%%%%%%%%%%%%%%%%%%%%%
\subsection{ When the initial drift is strictly negative $\mu < 0$  (case II)}
When the initial drift is strictly negative $\mu < 0$, the conditional probability density $\gamma(s\vert x,t)$ of the absorption at time $T_a$, knowing that the process started at the position $x$ at time $t$, is still given by~Eq.\eqref{gammafirstBrown} but with a non vanishing forever survival probability $S(\infty \vert x) = 1 -e^{2 \mu (a - x)}$~\cite{ref_book_Redner,ref_Monthus-Mazzolo}. Reporting these expressions in~Eq.\eqref{Qdef}, we obtain
\begin{align}
\label{Qmuneglambda}
	Q_{II}(x,t) & =
\int_t^{+\infty}   \frac{a }{\sqrt{2 \pi s^3} } e^{\lambda a - \frac{a^2}{2 s}- \frac{\lambda^2}{2}s}  \left(\frac{s}{s-t}  \right)^{\frac{3}{2} } \left( \frac{a-x}{a} \right)   
e^{- \mu x + \frac{\mu^2}{2} t +\frac{a^2}{2 s} - \frac{(a-x)^2}{2(s-t)}} ds \nonumber \\
			&+  (  1- e^{2 \lambda a } )  \left[ \frac{1- e^{2 \mu (a-x) } }{1- e^{2 \mu a } } \right]
\nonumber \\
%% 			&= \frac{(a-x ) }{\sqrt{2 \pi}  } e^{\lambda a- \mu x + \frac{\mu^2-\lambda^2}{2} t}
%% \int_t^{+\infty}    
%% \left(\frac{1}{s-t}  \right)^{\frac{3}{2} } 
%% e^{ - \frac{\lambda^2}{2} (s -t)  - \frac{(a-x)^2}{2(s-t)}} ds
%% +  (  1- e^{2 \lambda a } )  \left[ \frac{1- e^{2 \mu (a-x) } }{1- e^{2 \mu a } } \right]
%% \nonumber \\
%% 			&=\frac{(a-x ) }{\sqrt{2 \pi}  } e^{\lambda a- \mu x + \frac{\mu^2-\lambda^2}{2} t}
%% \frac{ \sqrt{2 \pi }}{ (a-x)} e^{  \lambda (a-x) }
%%  +\frac{\left(1-e^{2 a \lambda }\right)}{\left(1-e^{2 a \mu }\right)} \left(1-e^{2 \mu (a-x)}\right)
%% \nonumber \\
			&=  e^{2 a \lambda -\lambda  x-\mu  x +\frac{1}{2} \left(\mu ^2-\lambda ^2\right)t} +\frac{\left(1-e^{2 a \lambda }\right)}{\left(1-e^{2 a \mu }\right)} \left(1-e^{2 \mu (a-x)}\right) \, ,
\end{align}
and the corresponding drift
\begin{align}
\label{driftdoobmuneglambda}
 \mu^*_{II}(x,t) &  = \mu +  \partial_x \ln  Q_{II}(x,t) 
\nonumber  \\
			& =   \mu +  \partial_x \ln \left[ e^{2 a \lambda -\lambda  x-\mu  x +\frac{1}{2} \left(\mu ^2-\lambda ^2\right)t} +\frac{\left(1-e^{2 a \lambda }\right)}{\left(1-e^{2 a \mu }\right)} \left(1-e^{2 \mu (a-x)}\right)
\right] 
\nonumber  \\
			& = \mu + \frac{-(\lambda +\mu ) e^{2 a \lambda -\lambda  x-\mu  x +\frac{1}{2} \left(\mu ^2-\lambda ^2\right)t}+2 \mu \frac{\left(1-e^{2 a \lambda }\right)}{\left(1-e^{2 a \mu }\right)} e^{2 \mu (a-x)}}{e^{2 a \lambda -\lambda  x-\mu  x +\frac{1}{2} \left(\mu ^2-\lambda ^2\right)t} +\frac{\left(1-e^{2 a \lambda }\right)}{\left(1-e^{2 a \mu }\right)} \left(1-e^{2 \mu (a-x)}\right) } \, .
\end{align} 
Observe that when $\lambda = \mu$ then $\mu^*(x,t) = \mu$ as it should. However, as in the case of a positive initial drift, the stochastic representation of conditioned process, denoted as $X_{II}(t)$, is not that of a Brownian motion with drift $\lambda$ (again as one might have expected) but obeys a rather complex inhomogeneous diffusion process
\begin{align}
\label{SDEdriftdoobmuneglambda}
   dX_{II}(t)  = &\left(  \mu + \frac{-(\lambda +\mu ) e^{2 a \lambda -\lambda  X_{II}(t)-\mu  X_{II}(t) +\frac{1}{2} \left(\mu ^2-\lambda ^2\right)t}+2 \mu \frac{\left(1-e^{2 a \lambda }\right)}{\left(1-e^{2 a \mu }\right)} e^{2 \mu (a-X_{II}(t))}}{e^{2 a \lambda -\lambda  X_{II}(t)-\mu  X_{II}(t) +\frac{1}{2} \left(\mu ^2-\lambda ^2\right)t} +\frac{\left(1-e^{2 a \lambda }\right)}{\left(1-e^{2 a \mu }\right)} \left(1-e^{2 \mu (a-X_{II}(t))}\right) }  \right) dt \nonumber \\
              & + dW(t) \, , ~~ t \geq 0  \,  ,
\end{align} 
which depends on both the initial drift $\mu$ and the new drift $\lambda$. Assuming that $X_{II}(t)$ remains finite, the long-time behavior of the conditioned drift depends on the intensity of the two drifts, more precisely
\begin{eqnarray}
 \mu^*_{II}(x,t)   \underset{t \to \infty}{\sim\,} 
 \left\lbrace
  \begin{array}{lll}   
      -\mu \coth(\mu (a-x))
     &~~\mathrm{if~~} \vert \lambda \vert > \vert \mu \vert
    \\
      - \lambda  
     &~~\mathrm{if~~} \vert \lambda \vert < \vert \mu \vert
  \end{array}
\right.
\label{longtimemuII}
\end{eqnarray}
In the first case, $\vert \lambda \vert > \vert \mu \vert$, when $x$ approaches the boundary $a$, the conditioned drift behaves as
\begin{eqnarray}
	\mu^*_{II}(x,\infty)  \opsim_{x \to a^-}  -\frac{1}{a-x}  
\end{eqnarray}
which is the taboo drift we previously encountered, and consequently, the conditioned process can never cross the barrier $a$.
In the second case, when $\vert \lambda \vert < \vert \mu \vert$, the behavior becomes more surprising and counter-intuitive, since for large times the conditioned drift $-\lambda$ is positive instead of negative. We can guess that during the intermediate times the process has been pushed away from the barrier $a$; otherwise, a positive drift would lead to its absorption at the boundary $a$, which contradicts the original hypothesis of a positive survival probability. In the next paragraph, we will gain a better understanding of this weird behavior by calculating the probability densities of conditioned processes.

%%%%%%%%%%%%%%%%%%%%%%%%%%%%%%%%%%%%%%%%%%%%%%%%%%%%%%%%%%%%%%%%%%%%%%%%%%%%%%%%%%%%%%%%%%%%%%%%%%%%%%%%%%%%%%%%
%%																						%%
%%             STOCHASTIC DIFFERENTIAL EQUATION FOR AN ABSORBED BROWNIAN MOTION WITH DRIFT                    %%
%%																						%%
%%%%%%%%%%%%%%%%%%%%%%%%%%%%%%%%%%%%%%%%%%%%%%%%%%%%%%%%%%%%%%%%%%%%%%%%%%%%%%%%%%%%%%%%%%%%%%%%%%%%%%%%%%%%%%%%
\section{Probability densities of the conditioned processes}
\label{sec4}
In the previous paragraph, we left the reader in a puzzling yet interesting situation with two unexpected conditioned drifts. To gain a deeper comprehension of the behavior of these conditioned processes, we aim to determine their probability densities. 
One way for solving Eq.\eqref{SDEdriftdoobmuposlambda} or Eq.\eqref{SDEdriftdoobmuneglambda} is by Girsanov's transformation~\cite{ref_book_Karatzas,ref_book_Oksendal}. Let us be more specific and consider the Itô SDE
\begin{equation}
\label{def_BM_with_drift_SDE_2}
   \left\{
       \begin{aligned}
	  dX(t) & = \mu(X(t),t) dt +  dW(t)  \, , ~~ t \geq 0 \\       
	  X(0)  & = x_0 \, ,
       \end{aligned}
   \right.
\end{equation}
\noindent where $W(t)$ is a standard Brownian motion. Girsanov's theorem states that for any bounded measurable function $h$, the expectation of $h(X(t))$, where $X(t)$ is a solution of Eq.\eqref{def_BM_with_drift_SDE_2}, can be express as
\begin{equation}
\label{def_new_expectation}
	\E[h(X(t))] = \E[Z(t)h(\underbrace{x_0 + W(t)}_{\text{driftless process}})]  \, ,
\end{equation}
with
\begin{equation}
\label{def_Z}
   Z(t) =  e^{\textstyle\int_0^t \mu(x_0 + W(u),u) dW(u)  -\frac{1}{2} \int_0^t \mu(x_0 + W(u),u)^2 du}  \, .
\end{equation}
Moreover, the pair $(\tilde{X}(t),\tilde{W}(t))$ where 
\begin{eqnarray} 
 \left\lbrace
  \begin{array}{lll}
    \tilde{X}(t) = x_0 + W(t)
    \\
    \tilde{W}(t) = W(t) - \int_0^t \mu(x_0 + W(u)) du \, ,
  \end{array}
\right.  
\label{pait_Girsanov}
\end{eqnarray}
is a weak solution of Eq.\eqref{def_BM_with_drift_SDE_2} under the transformed probability density $\tilde{P}(\mathcal{X}(t)) = Z(t)P(\mathcal{X}(t))$, where $\mathcal{X}(t) = \{X(u) \vert \, 0 \leq u \leq t\}$ denotes the entire path of the Itô process $X(t)$ over the time interval $[0,t]$ (see Section 7.3 in~\cite{ref_book_Sarkka} for the proof and examples).

\noindent 
%Let us see what we get with the two drifts $\mu^*_{I}(x,t)$ and $\mu^*_{II}(x,t)$.
Now, let us examine what we get with the two drifts, $\mu^*_{I}(x,t)$ and $\mu^*_{II}(x,t)$.
%%%%%%%%%%%%%%%%%%%%%%%%%%%%%%%%%%%%%%%%%%%%%%%
%%       SUBSECTION: POSITIVE DRIFT  mu >= 0  %
%%%%%%%%%%%%%%%%%%%%%%%%%%%%%%%%%%%%%%%%%%%%%%%
\subsection{Probability density when the initial drift is vanishing or positive $\mu\geq 0$ (case I) }
The probability density $p_{I}(x,t)$ of the conditioned process defined by Eq.\eqref{SDEdriftdoobmuposlambda} is established in appendix~\ref{appendix_1} and can be expressed as follows:  
\begin{equation}
\label{PDFdriftdoobmuposlambda}
	p_{I}(x,t) = e^{2 a \lambda }  \left(e^{- \lambda x -\frac{1}{2} \lambda^2 t} +\left(e^{-2 a \lambda }-1\right) \frac{(a-x)}{a}\right) \frac{1}{\sqrt{2 \pi t}} \left(e^{-\frac{x^2}{2 t}}-e^{-\frac{(x-2 a)^2}{2 t}}\right),~t>0 . 
\end{equation}
From the previous equation it is straightforward to verify that $p_{I}(x,t)$ satisfies the Fokker-Planck equation
\begin{equation}
\label{Fokker-Planck-driftdoobmuposlambda}
	\partial_t p_{I}(x,t) -\frac{1}{2} \partial_{xx}^2  p_{I}(x,t) +\partial_{x}\left[\mu^*_{I}(x,t)  p_{I}(x,t) \right] = 0 \, .
\end{equation}
Besides, observe that the survival probability at time $t$
\begin{align}
\label{survival_at_time_t_muposlambdaneg}
	\int_{-\infty}^a p_{I}(x,t)  dx & = \int_{-\infty}^a e^{2 a \lambda }  \left(e^{- \lambda x -\frac{1}{2} \lambda^2 t} +\left(e^{-2 a \lambda }-1\right) \frac{(a-x)}{a}\right) \frac{1}{\sqrt{2 \pi t}} \left(e^{-\frac{x^2}{2 t}}-e^{-\frac{(x-2 a)^2}{2 t}}\right)  dx \nonumber \\
	                                & = \frac{1}{2} \left(1 + \text{erf}\left(\frac{a-\lambda  t}{\sqrt{2 t}}\right) - e^{2 a \lambda } \text{erfc}\left(\frac{a+\lambda  t}{\sqrt{2 t}}\right)\right) \, 
\end{align}
is that of the Brownian motion with drift $\lambda$, (Eq.\eqref{survival_at_time_t}), as it should.

%%%%%%%%%%%%%%%%%%%%%%%%%%%%%%%%%%%%%%%%%%%%%%%
%%       SUBSECTION: NEGATIVE DRIFT  mu < 0   %
%%%%%%%%%%%%%%%%%%%%%%%%%%%%%%%%%%%%%%%%%%%%%%%
\subsection{Probability density when the initial drift is strictly negative $\mu < 0$ (case II) }
\begin{figure}[h]
\centering
\includegraphics[width=5in,height=4.in]{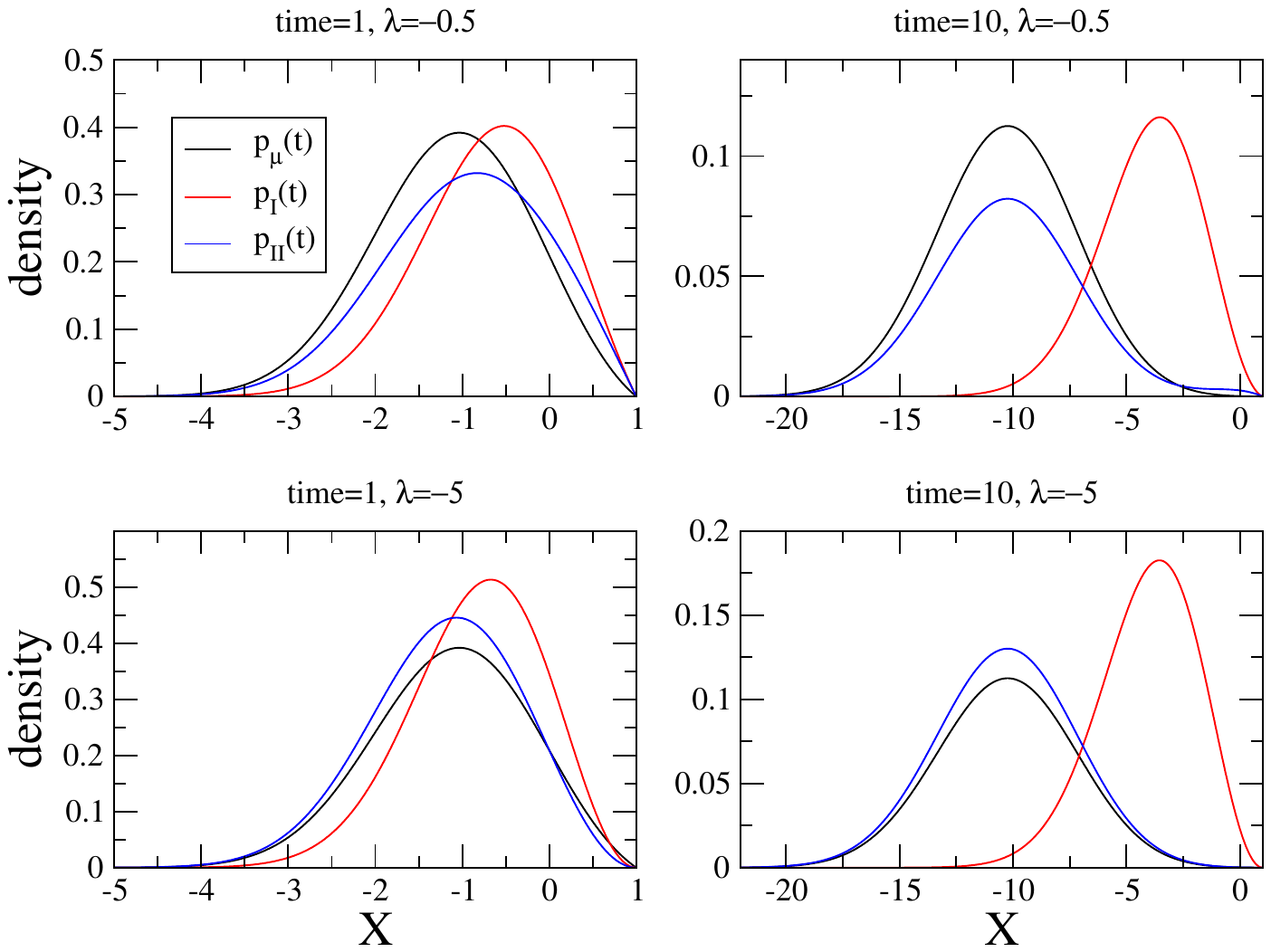}
\setlength{\abovecaptionskip}{15pt} 
\caption{Examples of density profiles for the three processes: $p_{\mu}(x,t)$ given by Eq.\eqref{imagesBrown} (black curve), $p_{I}(x,t)$ given by Eq.\eqref{PDFdriftdoobmuposlambda} (red curve) and $p_{II}(x,t)$ given by Eq.\eqref{PDFdriftdoobmuneglambda} (blue curve). 
The absorbing barrier is located at $a=1$. For the densities $p_{\mu}(x,t)$ and $p_{II}(x,t)$, the value of the parameter $\mu$ is $-1$. Consequently, the densities $p_{\mu}(x,t)$ are the same in each column, as they rely only on $\mu$ and the time $t$. 
In the top-right figure, observe the plateau in the density profile of $p_{II}(x,t)$ (blue curve) around $x\in[-2.5,0]$.}
\label{fig1}
\end{figure}

The probability density $p_{II}(x,t)$ of the conditioned process defined by Eq.\eqref{SDEdriftdoobmuneglambda} is derived in appendix~\ref{appendix_2} and is given by the expression
\begin{equation}
\label{PDFdriftdoobmuneglambda}
	p_{II}(x,t) =  \left(e^{2 a \lambda -\lambda  x-\mu  x +\frac{1}{2} \left(\mu ^2-\lambda ^2\right)t} +\frac{\left(1-e^{2 a \lambda }\right)}{\left(1-e^{2 a \mu }\right)} \left(1-e^{2 \mu (a-x)}\right) \right)
	  \frac{e^{\mu  x-\frac{1}{2}\mu ^2 t}}{\sqrt{2 \pi t}} \left(e^{-\frac{x^2}{2 t}}-e^{-\frac{(x-2 a)^2}{2 t}}\right), ~t>0 .
\end{equation}
From the previous equation, it is straightforward to verify that $p_{II}(x,t)$ satisfies the Fokker-Planck equation
\begin{equation}
\label{Fokker-Planck-driftdoobmuneglambda}
	\partial_t p_{II}(x,t) -\frac{1}{2} \partial_{xx}^2  p_{II}(x,t) +\partial_x\left[\mu^*_{II}(x,t)  p_{II}(x,t) \right] = 0 \, .
\end{equation}
Figure~\ref{fig1} shows various density profiles for the three processes $p_{\mu}(x,t)$, $p_{I}(x,t)$ and $p_{II}(x,t)$ at different times and for two different parameters $\lambda$.
%%%%%%%%%%%%%%%%%%%%%%%%%%%%%%%%%%%%%%%%%%%%%%%%%%%%%%%%%%%%%%%%%%%%%%%
%%                 FIGURE 1                    	        	 %
%%%%%%%%%%%%%%%%%%%%%%%%%%%%%%%%%%%%%%%%%%%%%%%%%%%%%%%%%%%%%%%%%%%%%%%                                  
%%%%%%%%%%%%%%%%%%%%%%%%%%%%%%%%%%%%%%%%%%%%%%%%%%%%%%%%%%%%%%%%%%%%%%%
It is worth noting that while the probability density $p_{II}(x,t)$ depends on both drifts $\mu$ and $\lambda$, it is remarkable that the survival probability at time $t$
\begin{align}
\label{survival_at_time_t_muneglambdaneg}
	\int_{-\infty}^a p_{II}(x,t)  dx & = \int_{-\infty}^a \left(e^{2 a \lambda -\lambda  x-\mu  x +\frac{1}{2} \left(\mu ^2-\lambda ^2\right)t} +\frac{\left(1-e^{2 a \lambda }\right)}{\left(1-e^{2 a \mu }\right)} \left(1-e^{2 \mu (a-x)}\right) \right) \nonumber \\ 
	                                 & \times \frac{e^{\mu  x-\frac{1}{2}\mu ^2 t}}{\sqrt{2 \pi t}} \left(e^{-\frac{x^2}{2 t}}-e^{-\frac{(x-2 a)^2}{2 t}}\right) dx \nonumber \\
%%	                                 & = \frac{1}{2} \left(1 + \text{erf}\left(\frac{a-\lambda  t}{\sqrt{2 t}}\right) - e^{2 a \lambda } \text{erfc}\left(\frac{a+\lambda  t}{\sqrt{2 t}}\right)\right)  \\
	                                 & = \Phi\left(a/\sqrt{t} - \lambda \sqrt{t} \right)  + e^{2 a \lambda } \left(1 - \Phi\left(a/\sqrt{t} + \lambda \sqrt{t} \right) \right) 
\end{align}
no longer depends on $\mu$, and coincides with that of the Brownian motion with $\lambda$ drift, (Eq. \eqref{survival_at_time_t}), as it should.

%%%%%%%%%%%%%%%%%%%%%%%%%%%%%%%%%%%%%%%%%%%%%%%
%%       SUBSECTION: AVERAGE BEHAVIOR         %
%%%%%%%%%%%%%%%%%%%%%%%%%%%%%%%%%%%%%%%%%%%%%%%
\subsection{Average behavior}
\begin{figure}[h]
\centering
\includegraphics[width=5in,height=4.in]{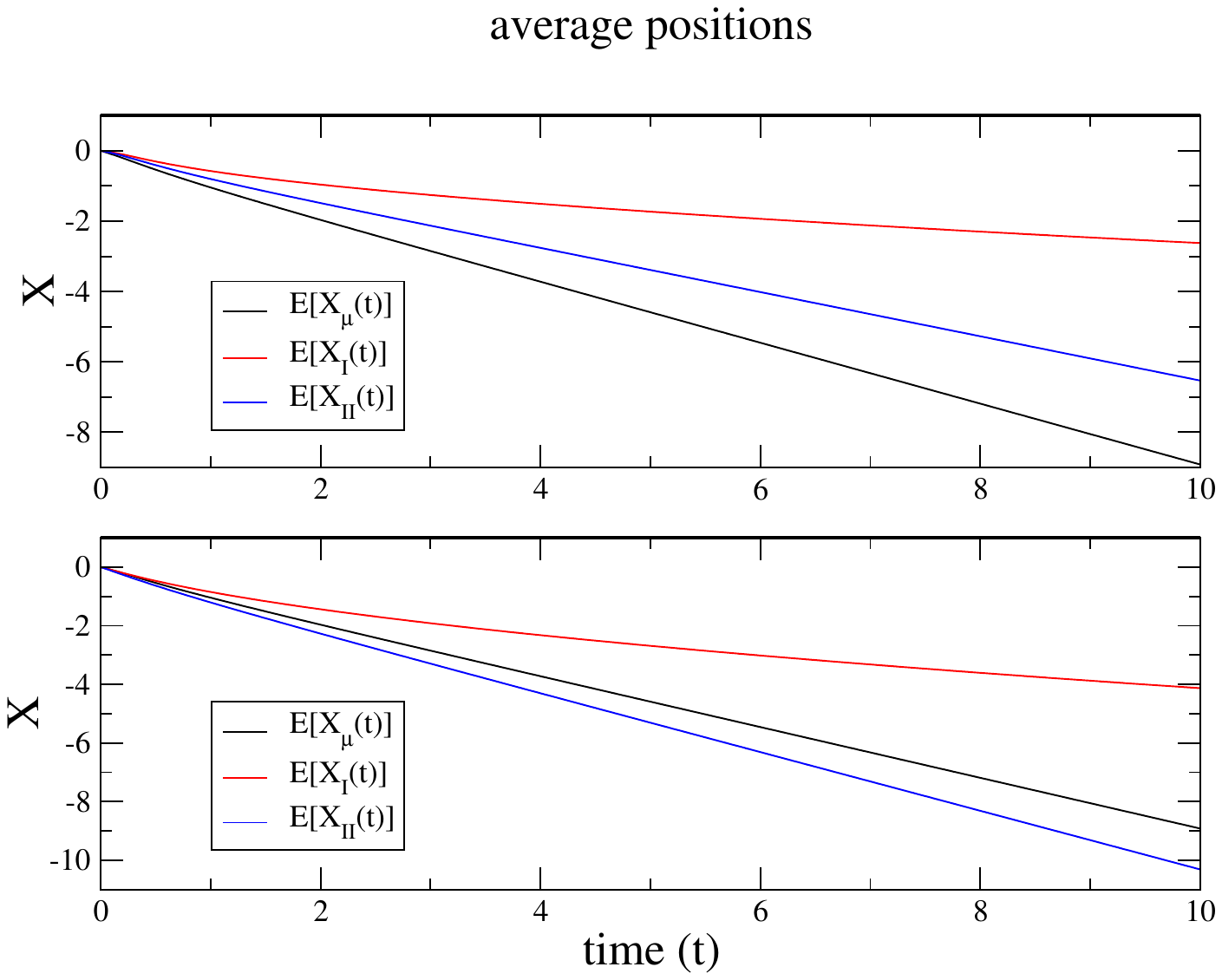}
\setlength{\abovecaptionskip}{15pt} 
\caption{Examples of average behaviors of the three processes: the process $X_{\mu}(t)$ given by Eq.\eqref{xmeanBrownian}, the type-I process given by Eq.\eqref{xmeantypeI} and the type-II process given by Eq.\eqref{xmeantypeII}. The absorbing barrier is located at $a=1$. Top figure $\mu = -1$ and $\lambda = -0.5$. Bottom figure $\mu = -1$ and $\lambda = -5$. Note that the black curve is the same in both figures since $\mu$ is identical.}
\label{fig2}
\end{figure}

From the transition densities, it is straightforward to obtain the mean behavior of the three processes. We obtain respectively
\begin{align}
\label{xmeanBrownian}
	\E[ X_{\mu}(t) ] = \int_{-\infty}^a x \, P_{\mu}(x,t)  dx & = \int_{-\infty}^a  x \, \frac{1}{\sqrt{2 \pi t}} \left( e^{- \frac{(x-\mu t )^2}{2t}} 
- e^{ 2 \mu a } e^{- \frac{(x-2a-\mu t)^2}{2t}} \right) dx \\
	                                 & = \frac{1}{2} \left(\mu  t \erf\left(\frac{a-\mu  t}{\sqrt{2} \sqrt{t}}\right)-e^{2 a \mu } (2 a+\mu  t) \erfc\left(\frac{a+\mu  t}{\sqrt{2} \sqrt{t}}\right)+\mu  t\right)  \nonumber
\end{align}
for the unconditioned Brownian motion with drift $\mu$ (where $\erf(x) = \frac{2}{\sqrt{\pi}} \int_0^{x} du \, e^{- u^2}$ is the Error function and $\erfc(x)$ the complementary Error function $\erfc(x) = 1 - \erf(x)$),
\begin{align}
\label{xmeantypeI}
	\E[ X_{I}(t) ] & = \int_{-\infty}^a x \, P_{I}(x,t)  dx = \int_{-\infty}^a  x \, e^{2 a \lambda }  \left(e^{- \lambda x -\frac{1}{2} \lambda^2 t} +\left(e^{-2 a \lambda }-1\right) \frac{(a-x)}{a}\right) \nonumber \\
	                                                                & \times \frac{1}{\sqrt{2 \pi t}} \left(e^{-\frac{x^2}{2 t}}-e^{-\frac{(x-2 a)^2}{2 t}}\right) dx \nonumber \\
	                                 & = \left(e^{2 a \lambda }-1\right) \left(a+\frac{t}{a}\right) \erf\left(\frac{a}{\sqrt{2 t}}\right) -2 a e^{2 a \lambda } +\sqrt{\frac{2 t}{\pi }}  \left(e^{2 a \lambda }-1\right) e^{-\frac{a^2}{2 t}} -\frac{1}{2} \lambda  t \left(e^{2 a \lambda }-1\right) \nonumber \\
	                                 & + \frac{1}{2} (2 a-\lambda  t) \erf\left(\frac{a-\lambda  t}{\sqrt{2 t}}\right) -\frac{1}{2} \lambda  t e^{2 a \lambda } \erf\left(\frac{a+\lambda  t}{\sqrt{2 t}}\right)
\end{align}
for the type-I process, and 
\begin{align}
\label{xmeantypeII}
	\E[  X_{II}(t) ] & = \int_{-\infty}^a x \, P_{II}(x,t)  dx = \int_{-\infty}^a  x \, \left(e^{2 a \lambda -\lambda  x-\mu  x +\frac{1}{2} \left(\mu ^2-\lambda ^2\right)t} +\frac{\left(1-e^{2 a \lambda }\right)}{\left(1-e^{2 a \mu }\right)} \left(1-e^{2 \mu (a-x)}\right) \right) \nonumber \\ 
	                                 & \times \frac{e^{\mu  x-\frac{1}{2}\mu ^2 t}}{\sqrt{2 \pi t}} \left(e^{-\frac{x^2}{2 t}}-e^{-\frac{(x-2 a)^2}{2 t}}\right) dx \nonumber \\
	                                 & = \frac{1}{4} (\coth (a \mu )-1)  \bigg[  \left(1 -e^{2 a \mu }\right) \left(e^{2 a \lambda } (2 a+\lambda  t)-\lambda  t\right) \nonumber \\
	                                 & + \lambda  t \left(e^{2 a \lambda }-e^{2 a (\lambda +\mu )}\right)\erf\left(\frac{a+\lambda  t}{\sqrt{2} \sqrt{t}}\right) +\left(e^{2 a \mu }-1\right) (2 a-\lambda  t) \erf\left(\frac{a-\lambda  t}{\sqrt{2} \sqrt{t}}\right)  \nonumber  \\
	                                 & +  2 \left(e^{2 a \lambda }-1\right) \left(e^{2 a \mu } (a+\mu  t) \erf\left(\frac{a+\mu  t}{\sqrt{2} \sqrt{t}}\right)+(\mu  t-a) \erf\left(\frac{a-\mu  t}{\sqrt{2} \sqrt{t}}\right)\right)  \bigg] 
\end{align}
for the type-II process.
For large times $t$, the leading order of the three previous expressions can be calculated
\begin{equation}
\label{xmeanasymptotic}
   \left\{
       \begin{aligned}
       \E[ X_{\mu}(t) ]  & \underset{t \to \infty}{\sim\,} \mu  \left(1 - e^{2 a \mu } \right) t  \\
       \E[ X_{I}(t) ]    & \underset{t \to \infty}{\sim\,}  -2 \sqrt{\frac{2}{\pi }}  \left(1- e^{2 a \lambda }\right)  \sqrt{t}\\
       \E[ X_{II}(t) ]   & \underset{t \to \infty}{\sim\,} \mu  \left(1 - e^{2 a \lambda } \right) t  \, .
       \end{aligned}
   \right.
\end{equation}
At large times, the average behavior of the unconditioned Brownian motion with drift $\mu$ and the average behavior of the type-II process behave like drifted Brownian motions with effective drift coefficients that are equal to $\mu  \left(1 - e^{2 a \mu } \right) $ and $\mu  \left(1 - e^{2 a \lambda } \right)$ respectively. 
%Observe that if $\vert \mu \vert > \vert \lambda \vert$, then at large times the average behavior of the type-II process  approaches $-\infty$ slower than the unconditioned Brownian motion with drift $\lambda$ as shown in Fig.~\ref{fig2}. In addition, due to the square-root term of time, at large times, the average behavior the type-I process stays closer to the boundary $a$ compared to the other two processes, no matter the values of $\mu$ and $\lambda$.
Observe that due to the square-root term of time, at large times, the average behavior the type-I process stays closer to the boundary $a$ compared to the other two processes, no matter the values of $\mu$ and $\lambda$, as shown in Fig.~\ref{fig2}.
%%%%%%%%%%%%%%%%%%%%%%%%%%%%%%%%%%%%%%%%%%%%%%%%%%%%%%%%%%%%%%%%%%%%%%%
%%                 FIGURE 2                    	        	 %
%%%%%%%%%%%%%%%%%%%%%%%%%%%%%%%%%%%%%%%%%%%%%%%%%%%%%%%%%%%%%%%%%%%%%%%                                  
%%%%%%%%%%%%%%%%%%%%%%%%%%%%%%%%%%%%%%%%%%%%%%%%%%%%%%%%%%%%%%%%%%%%%%%

%%%%%%%%%%%%%%%%%%%%%%%%%%%%%%%%%%%%%%%%%%%%%%%%%%%%%%%%%%%%%%%%%%%%%%%%%%%%%%%%%%%%%%%%%%%%%%%%%%%%%%%%%%%%%%%%
%%																						%%
%%                DISCUSSION                                                                                  %%
%%																						%%
%%%%%%%%%%%%%%%%%%%%%%%%%%%%%%%%%%%%%%%%%%%%%%%%%%%%%%%%%%%%%%%%%%%%%%%%%%%%%%%%%%%%%%%%%%%%%%%%%%%%%%%%%%%%%%%%
\section{Discussion}
\label{sec_Discussion}
In this article, we have considered the question: is the constant drift the only drift function such that for any $t>0$, the first-passage density to an absorbing boundary $a$ is given by $\gamma(t) = \frac{a}{\sqrt{2 \pi t^3} }  e^{-\frac{(a-\mu t)^2}{2 t}}$ ? Surprisingly, the answer is negative since we have discovered two families of diffusion processes, the first with one parameter (the type-I process) and the second with two parameters (the type-II process), both sharing the same first-passage density as that of a Brownian motion whose drift is directed away from the absorbing boundary.
Thus, while the "canonical process" (here the drifted Brownian motion) determines the first-passage density $\gamma(t)$, the converse is not generally true. 
This result also raises other questions: perhaps there are infinitely many processes all having $\gamma(t)$ as their first-passage density? Furthermore, does this property also hold for other processes, such as the Ornstein-Uhlenbeck process? 
It should be also considered that, for instance, the diffusion $Y(t) = W(t)^3$ driven by the SDE
\begin{equation}
	dY(t) = 3 Y(t)^{1/3} dt + 3 Y(t)^{2/3} dW(t),  ~~ Y(0)= 0 ,
\end{equation}
has the same FPT through the barrier $a=1$ as Brownian motion and therefore one finds that also the density of the FPT of $Y(t)$ through the barrier $a=1$ is the inverse Gaussian density with zero drift, i.e., $\gamma_0(t) = \frac{1}{\sqrt{2 \pi  t^3} }  e^{-\frac{1}{2 t}} \, , ~ t>0$. In other words, besides Brownian motion, the diffusion $Y(t)$ also has the inverse Gaussian density as FPT density, but only for a specific value of $a$ (of course, there are infinite straightforward examples of this kind). Finally, since the results presented in this article are rather unexpected, it would be very interesting to experimentally observe our predictions on real systems.\\

%%%%%%%%%%%%%%%%%%%%%%%%%%%%%%%%%%%%%%%%%%%%%%%%%%%%%%%%%%%%%%%%%%%%%%%%%%%%%%%%%%%%%%%%%%%%%%%%%%%%%%%%%%%%%%%%
%%																						%%
%%                 ACKNOWLEDGMENTS                                                                            %%
%%																						%%
%%%%%%%%%%%%%%%%%%%%%%%%%%%%%%%%%%%%%%%%%%%%%%%%%%%%%%%%%%%%%%%%%%%%%%%%%%%%%%%%%%%%%%%%%%%%%%%%%%%%%%%%%%%%%%%%
\section{Acknowledgements}
These two unexpected journeys are dedicated to the memory of Jérôme Combe.

\newpage
%%%%%%%%%%%%%%%%%%%%%%%%%%%%%%%%%%%%%%%%%%%%%%%%%%%%%%%%%%%%%%%%%%%%%%%%%%%%%%%%%%%%%%%%%%%%%%%%%%%%%%%%%%%%%%%%
%%																						%%
%%                 APPENDIX                                                                                   %%
%%																						%%
%%%%%%%%%%%%%%%%%%%%%%%%%%%%%%%%%%%%%%%%%%%%%%%%%%%%%%%%%%%%%%%%%%%%%%%%%%%%%%%%%%%%%%%%%%%%%%%%%%%%%%%%%%%%%%%%
\appendix

%%%%%%%%%%%%%%%%%%%%%%%%%%%%%%%%%%%%%%%%%%%%%%%%%%%%%%
%%                 APPENDIX 1                       %%
%%%%%%%%%%%%%%%%%%%%%%%%%%%%%%%%%%%%%%%%%%%%%%%%%%%%%%
\section{Probability density (case I) } 
\label{appendix_1}
In this appendix, we provide the derivation of the probability density for vanishing or positive initial drift  $\mu\geq 0$ (case I) using Girsanov's theorem, as detailed in the main text.\\

\noindent Reporting the expression of $\mu^*_{I}(x,t)$ into Eq.\eqref{def_Z} leads to (recall that $x_0=0$ in our cases)

\begin{align}
\label{ZtdriftI}
	Z_{I}(t) & =  e^{\textstyle \int_0^t \mu^*_{I}(W(u),u) dW(u)  -\frac{1}{2} \int_0^t \mu^*_{I}(W(u),u)^2 du}  \nonumber \\
	     & =  e^{\textstyle \int_0^t \left(\partial_x \log[\tilde{Q}_{I}(W(u),u)] \right) dW(u)  -\frac{1}{2} \int_0^t \left(\partial_x \log[\tilde{Q}_{I}(W(u),u)] \right)^2 du}  \nonumber \\
	     & =  e^{\textstyle \int_0^t  \frac{\partial_x \tilde{Q}_{I}(W(u),u)}{\tilde{Q}_{I}(W(u),u)} dW(u)   -\frac{1}{2} \int_0^t  \frac{(\partial_x \tilde{Q}_{I}(W(u),u))^2}{\tilde{Q}_{I}(W(u),u)^2} du }
\end{align}
with (Eq.\eqref{driftdoobmuposlambda})
\begin{equation}
   \left\{
       \begin{aligned}
       \tilde{Q}_{I}(x,t)  & = e^{  - \lambda x - \frac{\lambda^2}{2}t } + (   e^{-2 \lambda a } -1)   \left(  \frac{a-x }{a}   \right)   \\
       \tilde{Q}_{I}(0,0)  & = e^{-2 \lambda a } \, .
       \end{aligned}
   \right.
\end{equation}
The first integral can be evaluated thanks to Itô's formula,

\begin{align}
\label{itoformulaQI}
d\left[\log \left[\tilde{Q}_{I}(W(t),t)\right]  \right] & =	d\left[\log \left[e^{  - \lambda W(t) - \frac{\lambda^2}{2}t } + (   e^{-2 \lambda a } -1)   \left(  \frac{a-W(t) }{a}   \right) \right] \right]  \nonumber \\
		& = \cancel{ -\frac{\lambda^2}{2} \frac{e^{-\lambda W(t) - \frac{\lambda^2}{2}t} }{\tilde{Q}_{I}(W(t),t)} }dt + \frac{\partial_x \tilde{Q}_{I}(W(t),t)}{\tilde{Q}_{I}(W(t),t)} dW(t)  \nonumber \\
		& +\frac{1}{2}\left[\cancel{  \frac{\lambda^2 e^{-\lambda W(t) - \frac{\lambda^2}{2}t} }{\tilde{Q}_{I}(W(t),t)} } - \frac{(\partial_x \tilde{Q}_{I}(W(t),t))^2}{\tilde{Q}_{I}(W(t),t)^2}  \right] dt  \, .
\end{align}
Therefore, the first integral reads
\begin{align}
\label{First_integral_driftdoobmuposlambda}
	\int_0^t  \frac{\partial_x \tilde{Q}_{I}(W(u),u)}{\tilde{Q}_{I}(W(u),u)} dW(u)  = \log \left[\tilde{Q}_{I}(W(t),t)\right] - \log \left[\tilde{Q}_{I}(0,0)\right] + \frac{1}{2} \int_0^t  \frac{(\partial_x \tilde{Q}_{I}(W(u),u))^2}{\tilde{Q}_{I}(W(u),u)^2} du  \, .
\end{align}
Reporting this expression into Eq.\eqref{ZtdriftI}, we obtain
\begin{align}
\label{ZtdriftIfinal}
	Z_{I}(t) & =  e^{\textstyle\log \left[\tilde{Q}_{I}(W(t),t)\right] - \log \left[\tilde{Q}_{I}(0,0)\right]} 
	 \nonumber \\
			 & = e^{2 \lambda a } \left[ e^{  - \lambda W(t) - \frac{\lambda^2}{2}t } + (   e^{-2 \lambda a } -1)   \left(  \frac{a-W(t) }{a}   \right) \right]  \, .
\end{align}
Since this expression depends solely on the state of the Brownian motion at time $t$, the probability density can be explicitly evaluated~\cite{ref_book_Sarkka}. Recall that the probability density of the drift-less process $\tilde{X}(t) = x_0 + W(t) = W(t)$ is given by~Eq.\eqref{imagesBrown} with $\mu =0$
\begin{equation}
\label{imagesBrowndriftless}
	 P_{0}(\tilde{X},t)  = \frac{1}{\sqrt{2 \pi t}} \left( e^{- \frac{\tilde{X}^2}{2t}} 
- e^{- \frac{(\tilde{X}-2a)^2}{2t}} \right),~t>0,
\end{equation}
and thus from Eqs.\eqref{ZtdriftIfinal} and \eqref{imagesBrowndriftless} we obtain the probability density $p_{I}(x,t)$ of the conditioned process Eq.\eqref{SDEdriftdoobmuposlambda}
\begin{equation}
\label{PDFdriftdoobmuposlambda_Appendice}
	p_{I}(x,t) = e^{2 a \lambda }  \left(e^{- \lambda x -\frac{1}{2} \lambda^2 t} +\left(e^{-2 a \lambda }-1\right) \frac{(a-x)}{a}\right) \frac{1}{\sqrt{2 \pi t}} \left(e^{-\frac{x^2}{2 t}}-e^{-\frac{(x-2 a)^2}{2 t}}\right),~t>0, 
\end{equation}
which, by construction, fulfills the boundary condition
\begin{equation}
\label{PDF_BC_driftdoobmuposlambda}
	p_{I}(a,t) = 0  \, .
\end{equation}

%%%%%%%%%%%%%%%%%%%%%%%%%%%%%%%%%%%%%%%%%%%%%%%%%%%%%%
%%                 APPENDIX 2                       %%
%%%%%%%%%%%%%%%%%%%%%%%%%%%%%%%%%%%%%%%%%%%%%%%%%%%%%%
\section{Probability density (case II) } 
\label{appendix_2}
In this appendix, we give the derivation of the probability density when the initial drift is negative $\mu < 0$ (case II) thanks to Girsanov's theorem.\\

\noindent As in the case I, we proceed by inserting the expression of $\mu^*_{II}(x,t)$ in Eq.\eqref{def_Z}, resulting in
\begin{align}
\label{ZtdriftII}
	Z_{II}(t) & =  e^{\textstyle \int_0^t \mu^*_{II}(W(u),u) dW(u)  -\frac{1}{2} \int_0^t \mu^*_{II}(W(u),u)^2 du}  \nonumber \\
	     & =  e^{\textstyle \int_0^t \left( \mu + \partial_x \log[Q_{II}(W(u),u)] \right) dW(u)  -\frac{1}{2} \int_0^t \left( \mu + \partial_x \log[Q_{II}(W(u),u)] \right)^2 du}  \\
	     & =  e^{\textstyle \mu W(t) + \int_0^t  \frac{\partial_x Q_{II}(W(u),u)}{Q_{II}(W(u),u)} dW(u) -\frac{1}{2} \mu^2 t - \mu \int_0^t  \frac{\partial_x Q_{II}(W(u),u)}{Q_{II}(W(u),u)} du  -\frac{1}{2} \int_0^t  \frac{(\partial_x Q_{II}(W(u),u))^2}{Q_{II}(W(u),u)^2} du }  \nonumber
\end{align}
with (Eq.\eqref{driftdoobmuneglambda})
\begin{equation}
   \left\{
       \begin{aligned}
       Q_{II}(x,t)      & = e^{2 a \lambda -\lambda  x-\mu  x +\frac{1}{2} \left(\mu ^2-\lambda ^2\right)t} +\frac{\left(1-e^{2 a \lambda }\right)}{\left(1-e^{2 a \mu }\right)} \left(1-e^{2 \mu (a-x)}\right)  \\
       Q_{II}(0,0)  & = 1 \, .
       \end{aligned}
   \right.
\end{equation}
The stochastic integral can be evaluated by applying the Itô formula
\begin{align}
\label{itoformulaQII}
	d\left[\log \left[Q_{II}(W(t),t)\right]  \right] & =	\frac{\partial_t Q_{II}(W(t),t)} {Q_{II}(W(t),t)} dt +
     \frac{1}{2} \left[ \frac{\partial_{xx}^2 Q_{II}(W(t),t)} {Q_{II}(W(t),t)} - \frac{(\partial_x Q_{II}(W(t),t))^2} {Q^2_{II}(W(t),t)} \right] dt \nonumber \\
	 & + \frac{\partial_x Q_{II}(W(t),t)} {Q_{II}(W(t),t)} dW(t) 
\end{align}
which reads,
\begin{align}
\label{First_intergral_driftdoobmuneglambda}
	\int_0^t  \frac{\partial_x Q_{II}(W(u),u)}{Q_{II}(W(u),u)} dW(u) & = \log \left[Q_{II}(W(t),t)\right] - \cancel{\log \left[Q_{II}(0,0)\right]} + \frac{1}{2} \int_0^t  \frac{(\partial_x Q_{II}(W(u),u))^2}{Q_{II}(W(u),u)^2} du  \nonumber \\
				& - \int_0^t \frac{\partial_t Q_{II}(W(u),u)} {Q_{II}(W(u),u)} du -  \frac{1}{2} \int_0^t  \frac{\partial_{xx}^2 Q_{II}(W(u),u)} {Q_{II}(W(u),u)} du  \, ,
\end{align}
and Eq.\eqref{ZtdriftII} becomes
\begin{align}
\label{ZtdriftIIfinal}
	Z_{II}(t) & = e^{\textstyle \mu W(t) + \log \left[Q_{II}(W(t),t)\right]- \int_0^t \frac{\partial_t Q_{II}(W(u),u)} {Q_{II}(W(u),u)} du - \frac{1}{2} \int_0^t  \frac{\partial_{xx}^2 Q_{II}(W(u),u)} {Q_{II}(W(u),u)} du + \cancel{\frac{1}{2} \int_0^t  \frac{(\partial_x Q_{II}(W(u),u))^2}{Q_{II}(W(u),u)^2} du} }  \nonumber \\
	     & \times  e^{\textstyle -\frac{1}{2} \mu^2 t - \mu \int_0^t  \frac{\partial_x Q_{II}(W(u),u)}{Q_{II}(W(u),u)} du  \cancel{-\frac{1}{2} \int_0^t  \frac{(\partial_x Q_{II}(W(u),u))^2}{Q_{II}(W(u),u)^2} du} }  \nonumber  \\
	     & = Q_{II}(W(t),t) e^{\textstyle \mu W(t) -\frac{1}{2} \mu^2 t} e^{ - \int_0^t \overbrace{ \left( \scriptstyle \frac{\partial_t Q_{II}(W(u),u)} {Q_{II}(W(u),u)} +\mu \frac{\partial_x Q_{II}(W(u),u)}{Q_{II}(W(u),u)} + \frac{1}{2} \frac{\partial_{xx}^2 Q_{II}(W(u),u)} {Q_{II}(W(u),u)} \right)}^{=0 \text{ by virtue of Eq.}\eqref{Qbackward} } du }
	    \nonumber  \\
	     & =  \left( e^{2 a \lambda -\lambda  W(t)-\mu  W(t) +\frac{1}{2} \left(\mu ^2-\lambda ^2\right)t} +\frac{\left(1-e^{2 a \lambda }\right)}{\left(1-e^{2 a \mu }\right)} \left(1-e^{2 \mu (a-W(t))}\right) \right)	 e^{\textstyle \mu W(t) -\frac{1}{2} \mu^2 t}     
\end{align}
As in case I, this expression depends only on the state of the Brownian motion at time $t$. Thus, we can explicitly evaluate the probability density $p_{II}(x,t)$ of the conditioned process, and we obtain
\begin{equation}
\label{PDFdriftdoobmuneglambda_Appendice}
	p_{II}(x,t) =  \left(e^{2 a \lambda -\lambda  x-\mu  x +\frac{1}{2} \left(\mu ^2-\lambda ^2\right)t} +\frac{\left(1-e^{2 a \lambda }\right)}{\left(1-e^{2 a \mu }\right)} \left(1-e^{2 \mu (a-x)}\right) \right)
	  \frac{e^{\mu  x-\frac{1}{2}\mu ^2 t}}{\sqrt{2 \pi t}} \left(e^{-\frac{x^2}{2 t}}-e^{-\frac{(x-2 a)^2}{2 t}}\right),~t>0,
\end{equation}
which, by construction, satisfies the boundary condition
\begin{equation}
\label{PDF_BC_driftdoobmuneglambda}
	p_{II}(a,t) = 0  \, .
\end{equation}

\newpage
%%%%%%%%%%%%%%%%%%%%%%%%%%%%%%%%%%%%%%%%%%%%%%%%%%%%%%%%%%%%%%%%%%%%%%%%%%%%%%%%%%%%%%%%%%%%%%%%%%%%%%%%%%%%%%%%
%%																						%%
%%                BIBLIOGRAPHY                                                                                %%
%%																						%%
%%%%%%%%%%%%%%%%%%%%%%%%%%%%%%%%%%%%%%%%%%%%%%%%%%%%%%%%%%%%%%%%%%%%%%%%%%%%%%%%%%%%%%%%%%%%%%%%%%%%%%%%%%%%%%%%

\end{document}